\documentclass[aip,jcp,amsmath,amssymb]{revtex4-1}

\usepackage[english]{babel}
\usepackage[applemac]{inputenc}
\usepackage{cmap}
\usepackage[T1]{fontenc}
\usepackage{textcomp}
\usepackage{graphicx}
\usepackage{dcolumn}
\usepackage{bm}
\usepackage{amsmath}
\usepackage{xcolor}

\newcommand\e[1]{\ensuremath{_{\text{#1}}}}
\newcommand\ex[1]{\ensuremath{^{\text{#1}}}}

\begin{document}

\title{Free energy landscapes for the thermodynamic understanding of adsorption-induced deformations and structural transitions in porous materials}

\author{D. Bousquet}
\affiliation{CNRS-ENS-UPMC, D\'epartement de Chimie, \'Ecole Normale Sup\'erieure, 24 rue Lhomond, 75005 Paris, France}
\author{F.-X. Coudert}
\affiliation{CNRS \& Chimie ParisTech, 11 rue Pierre et Marie Curie, 75005 Paris, France}
\author{A. Boutin}
\email{anne.boutin@ens.fr}
\affiliation{CNRS-ENS-UPMC, D\'epartement de Chimie, \'Ecole Normale Sup\'erieure, 24 rue Lhomond, 75005 Paris, France}

\date{\today}

\begin{abstract}

Soft Porous Crystals are flexible metal--organic frameworks that respond to physical stimuli such as temperature, pressure and gas adsorption by large changes in their structure and unit cell volume. While they have attracted a lot of interest, molecular simulation methods that directly couple adsorption and large structural deformations in an efficient manner are still lacking. We propose here a new Monte Carlo simulation method based on non-Boltzmann sampling in \{guest loading, volume\} space using the Wang--Landau algorithm, and show that it can be used to fully characterize the adsorption properties and the material's response to adsorption at thermodynamic equilibrium. We showcase this new method on a simple model of the MIL-53 family of breathing materials, demonstrating its potential and contrasting it with the pitfalls of direct, Boltzmann simulations. We furthermore propose an explanation for the hysteretic nature of adsorption in terms of free energy barriers between the two metastable host phases.

\end{abstract}



\maketitle

\section{Introduction}


A new field of research has emerged in the past decade in the context of solid-state chemistry and physical chemistry. It is the science of Metal--Organic Frameworks (MOF's). Metal--Organic Frameworks, also called \emph{Porous Coordination Polymers} (PCP's), are hybrid crystalline porous materials consisting of metallic species connected to one another with organic linkers. They display an extremely large range of crystal structures and host-guest properties, which makes them an important class of materials with potentially major impact in adsorption/separation technologies of strategic gas linked with energy supply and environmental problems. The combination of tunable porosity, the functionalization of the internal surface together with the structural flexibility of the host opens the way to an extremely rich host-guest chemistry, putting this class of materials in a unique position.

One fascinating aspect of hybrid frameworks is the ability of a subclass of structures to behave in a remarkable guest-responsive fashion. These so-called \emph{Soft Porous Crystals} (SPC's)\cite{Kit2009} exhibit a variety of large amplitude dynamic behaviors of their frameworks in response to external stimuli of weak intensity (light, electric field, gas exposure, etc.). The change in the SPC channels in response to the external constraint is reversible and maintains the crystalline character of the solid. As an example, one may cite the MIL-53 type frameworks\cite{Ser2002,Ser2007} which exhibit guest-induced structural phase transitions upon gas adsorption and desorption, called ``breathing'' transitions. The bistable behavior of this system is controlled by the gas pressure that acts as the external stimulus.


Molecular simulation of adsorption in flexible porous solids is a challenging field,\cite{Cou2011} and several approaches have been developed over the time to circumvent the difficulty of direct simulation in the osmotic ensemble, in which the temperature $T$, mechanical strain $\sigma$, host framework $N\e{host}$ and adsorbate chemical potential $\mu\e{ads}$ are imposed. Firstly, many published works trying to shed light onto the interplay between adsorption and structural changes have relied on an indirect approach, in which regular Grand Canonical Monte Carlo (GCMC) simulations (i.e. at constant chemical potential, temperature and volume) are performed on a number of possible host structures. Adsorption isotherms computed using this approach are then compared with experimental data, and any steps observed between ``rigid host'' isotherms can be ascribed to adsorption-induced structural transitions. This was successfully used to demonstrate that hydrogen adsorption could trigger flexibility in ZIF-8,\cite{Dur2011} as well as in early simulation studies of the breathing of MIL-53(Cr) upon CO$_2$ adsorption.\cite{Ram2007} While in most cases these GCMC simulations were performed on rigid adsorbents, some authors have used a forcefield accounting for the flexibility of the framework, describing local deformations of the host material even though the overall unit cell volume is preserved.\cite{Mau2011}

A second approach to guest-induced flexibility of soft porous crystals is to perform a series of $(N\e{ads},P,T)$ molecular simulations at increasing loading $N\e{ads}$. Dubbeldam et al. have showcased a methodology \cite{Dub2009} where the material's unit cell, loaded with adsorbate, is energy-minimized. This allows one to follow the zero-Kelvin response of the material's structure to guest adsorption, and describes elastic deformation (generated by the so-called adsorption stress) as well as structural transitions. Others groups have performed $(N\e{ads},\sigma,T)$ 
simulation either without adsorbate \cite{Yot2012new} or with a varying number of guest molecules,\cite{Ma2012,Sal2008new,Gho2010new} 
as in the case of H$_2$O in MIL-53(Cr).\cite{Sal2011}


Compared to these indirect approaches, studies using direct molecular simulation in the osmotic ensemble are scarce. Maurin and coworkers \cite{Gho2010} have reported the use of an hybrid Molecular Dynamics/Monte Carlo scheme (MD/MC) \cite{Dua1987} for the description of CO$_2$ adsorption in MIL-53(Cr). This scheme includes, as part of a $(\mu\e{ads},V,T)$ Monte Carlo simulation, shorts runs of $(N\e{ads},P,T)$ molecular dynamics in which the unit cell of the system can deform. These authors demonstrated that such a scheme presents severely limiting problems of convergence towards equilibrium, resulting in only some of the structural transitions being reproduced during an adsorption--desorption cycle. Recently, those authors have shown that a careful and thorough calibration of force field parameters could enable them to witness the second transition during adsorption\cite{ASAPnew}. In essence, direct simulation in the osmotic ensemble 
does not easily allow one to efficiently overcome free energy barriers between very different states of the system. It can, however, be successful if the different host structures have similar energy and volume, as is the case of silicalite-1, where the three possible phases of the material differ only by 0.6\% in unit cell volume.\cite{Jef2008}

Finally, a last class of simulation methods for adsorption in flexible nanoporous materials relies on the use of free energy methods (also called density of state calculations). These techniques are based on the computation of the full free energy landscape as a function of one or more order parameters. The thermodynamic behavior of the system can then be fully determined from this landscape. For example, Miyahara et al. studied the adsorption-induced structural transitions of a so-called jungle gym model system through the reconstruction of grand free energy profiles by thermodynamic integration as a function of subnet displacement.\cite{Wat2009,Sug2012} By choosing both the unit cell loading and cell vectors as order parameters, one can reconstruct free energy landscapes directly from adsorption isotherms in a rigid host. Similar reconstruction has been successfully applied to the MIL-53 materials by an analytical approach that uses fits of experimental isotherms.\cite{Cou2008,Bou2009,Bou2010}

In this work, we extend these ideas and present a new Monte Carlo simulation method using a non-Boltzmann sampling algorithm to perform direct molecular simulation of adsorption in flexible nanoporous solids. This method is compared to standard GCMC calculation and direct osmotic simulation.

\section{Theory}

\subsection{The Osmotic Ensemble}

The adequate ensemble to study adsorption in flexible materials is the osmotic ensemble introduced by Brennan and Madden\cite{Bre2002} in the context of polymer-solvent mixtures. The control parameters for this ensemble are the temperature, $T$, the mechanical constraint exerted on the material, $\sigma$, the chemical potential of the adsorbed gas, $\mu\e{ads}$, and the number of  particles of the host framework, $N\e{host}$. The osmotic grand potential is written as follows:
\begin{equation}
\Omega\e{os}(N\e{host},\mu\e{ads},\sigma,T) = U - TS + \sigma V - \mu\e{ads} N\e{ads}
\end{equation}
\noindent If the host exhibits structural phase transitions, i.e. if it exists in an equilibrium between a number of metastable structures, the host degrees of freedom may be decoupled from the rest of the systems variables. One can then decompose the osmotic grand potential as a sum of the free energy of the host material and the grand canonical potential for the guest molecules.\cite{Cou2011} This was used extensively to study the MIL-53(Al) adsorption induced breathing, mainly through an analytical scheme.\cite{Cou2008,Bou2009} The osmotic potential is then written as:
\begin{equation}
\Omega\e{os}(N\e{host},\mu\e{ads},\sigma,T) = F\e{host}(V,N\e{host},T) +\sigma V + \Omega\e{GC}(\mu\e{ads},T;V)
\end{equation}
\noindent Theoretical models based on this decomposition of the osmotic potential have met success in allowing the interpretation of data obtained from either adsorption--desorption experiments\cite{Cou2008,Bou2010,Bou2009} or molecular simulation in the grand canonical ensemble.\cite{Jef2008} However, direct molecular simulation is constrained by severe fundamental limitations, which were highlighted in the previous section. Both Monte Carlo and molecular dynamics-based techniques typically cannot overcome free energy barriers coupling adsorption and host deformation, unless the host phases are very close energetically and geometrically. Here we present a molecular simulation methodology able to bypass these limitations by relying on non-Boltzmann sampling based on the use of a Wang--Landau algorithm.

\subsection{The Wang--Landau algorithm}

\noindent The Wang--Landau algorithm was originally developed to calculate density of states on the fly in the canonical ensemble,\cite{Wang2001} by performing a Monte Carlo simulation with a non-Boltzmann acceptance probability of $\mathcal P(E_1\rightarrow E_2) = \min\left(\frac{g(E_1)}{g(E_2)},1\right)$, where $g(E)$ is the density of states of the system. This technique was later extended to other ensembles and the calculation of free energy as a function of other order parameters (energy, volume, number of particles, reaction coordinate, etc.).\cite{Pab2002, Pab2004, Pab2006, Zho2005, Cai2008, Cai2009} We propose here that the Wang--Landau can be applied to the calculation of the free energy in the osmotic ensemble, as a function of the order parameters involved in the adsorption-induced structural transitions: the unit cell loading $N\e{ads}$ (i.e. the quantity of adsorbed particles) and the host unit cell volume $V$.  The density of state also depends on temperature $T$ and number of particles of the host framework, $N\e{host}$. Since these thermodynamics variables are fixed in all our study, there are omitted in our notation in the following. The knowledge of the extended density of states $Q(N\e{ads},V)$ of a given system allows to calculate all of its thermodynamic properties. The osmotic thermodynamic potential can thus be expressed as:

\begin{equation}
\Omega\e{os} (\mu\e{ads},\sigma) = - k_B T \ln \left[ \sum_{V,N\e{ads}} Q(N\e{ads},V) \,e^{\beta(\mu\e{ads} N\e{ads} - \sigma V)}\right]
\end{equation}

\noindent while the osmotic Wang--Landau free energy for a given chemical potential $\mu\e{ads}$ and external pressure $\sigma$ is: 

\begin{equation}
\Omega\e{WL} (N\e{ads},V;\mu\e{ads},\sigma) = -k_B T \ln \left[ Q(N\e{ads},V) e^{\beta(\mu\e{ads} N\e{ads}-\sigma V)} \right]
\end{equation}

\noindent Thus, once the two-dimensional extended density of states for the system has been calculated for a given host and a fixed temperature, the osmotic potential as well as the Landau free energy can be derived from this density of state for any set of parameters $(\mu\e{ads},\sigma)$. In turn, any observable may then be similarly computed \emph{a posteriori}, as a function of chemical potential and mechanical pressure, from a single extended density of state calculation.

We now turn to the practical calculation of this two-dimensional density of states during a Monte Carlo simulation. During a ``standard'' Monte Carlo simulation using Boltzmann acceptance probabilities, the probability to generate a configuration with $N\e{ads}$ adsorbed particles and a volume $V$ is proportional to $Q(N\e{ads},V)$. The idea behind the Wang--Landau algorithm is to modify the acceptance probabilities of the insertion/deletion and volume change MC moves so that the simulation will homogeneously visit all states. This means that acceptance probabilities are multiplied by the ratio of the density of state between the old and new states:
\begin{eqnarray}
\mathcal P(N\e{ads},V\e{old}\rightarrow N\e{ads},V\e{new}) &=& \textrm{min} \left[ 1,
\frac{Q(N\e{ads},V\e{old})}{Q(N\e{ads},V\e{new}) }
\left(\frac{V\e{new}}{V\e{old}} \right)^{N\e{ads}}
\textrm{e}^{-\beta \Delta E}  \right]\\[1ex]
\mathcal P(N\e{ads},V\rightarrow N\e{ads}+1,V) &=& \textrm{min} \left[ 1,
\frac{Q(N\e{ads},V)}{Q(N\e{ads}+1,V)}\cdot
\frac{V}{(N\e{ads}+1)\lambda^3}
\textrm{e}^{-\beta \Delta E} \right]\\[1ex]
\mathcal P(N\e{ads},V\rightarrow N\e{ads}-1,V) &=& \textrm{min} \left[ 1,
\frac{Q(N\e{ads},V)}{Q(N\e{ads}-1,V)}\cdot
\frac{N\e{ads}\lambda^3}{V}
\textrm{e}^{-\beta \Delta E} \right]
\end{eqnarray}

\noindent where $\lambda$ is the de Broglie wavelength. The density of state (DOS) $Q(N\e{ads},V)$ is discretized and updated on the fly during the Monte Carlo simulation. The volume $V$ is a continuous variable, so it is discretized into a series of bins $[V_i;V_i+\delta_i V]$; for each value of $N\e{ads}$ (which is naturally discrete) and each bin of $V$, the computer stores and updates the value of $Q$. Because it is initially unknown, it is set uniformly to 1 at the beginning of the simulation. Then, at each Monte Carlo step, the DOS of the current $(N\e{ads},V)$ state is multiplied by a factor $f$. Simulation proceeds until all states have been uniformly visited, which is indicated by a flat histogram of visited states in $(N\e{ads},V)$ space. When the histogram is flat, the accumulated density of state has converged with an accuracy proportional to $\ln(f)$. Then, the histogram is cleared, the factor $f$ is decreased (we used the $f \rightarrow \sqrt{f}$ scheme in this study), and the simulation is allowed to continue for a new Wang--Landau iteration, with this new value of $f$.

Convergence of the DOS depends on the overall number of Wang--Landau iterations performed and the convergence of the DOS in the course of each iteration. To assess the flatness of the histogram, which determines the latter, one of two schemes is typically used. First, one can either consider a minimum visit criterium, i.e. consider the histogram is flat when all bins were at least visited a specific number of times (we chose 500/$\sqrt{\ln f}$ in this study). A second criterion is to require that the least visited and most visited bins of the DOS are within 20\% of the mean number of visits. Both approaches yielded the same result on our system, and we there used first requirement (minimum number of visits per bin), which is less demanding in terms of computational cost.\cite{Zho2005} Furthermore, $f$ has to be initially high enough to allow the system to rapidly visit all states, so we chose its initial value as $e^4$. The Wang--Landau free energy profiles obtained thusly converged after 23~Wang--Landau iterations, with a final value of $f \simeq \exp(10^{-6}) \simeq 1+10^{-6}$.

\subsection{Reduction of the Two-Dimensional Parameter Space}

The two-dimensional Wang--Landau sampling in $(N\e{ads},V)$ space as outlined above forces a broad sampling of the system's configurational space, overcoming free energy barriers coupling adsorption and framework deformation. However, the dimensionality of the order parameter space and the stringent conditions imposed on the convergence of the density of state mean the methods has a very expensive computational cost. One possible approach to reduce the CPU cost is that the problem is \emph{embarrassingly parallel} in nature: independent simulations can be run on disjunct sets of the order parameter space, and the full density of state later reconstructed from all simulations by continuity. But such lengthy calculations\footnote{We performed such simulations on a commodity cluster of 4 Intel Xeon CPU cores at 3.0~GHz, but found that simulation time for full convergence exceeded practical limits; after 30 days of simulation on the cluster, only 10 Wang--Landau iterations had been performed.} are not in fact necessary, as other schemes can be used to reconstruct a full 2D free energy profile from 1D Wang--Landau calculations. 

\begin{figure}[ht]
\includegraphics[width=0.6\textwidth]{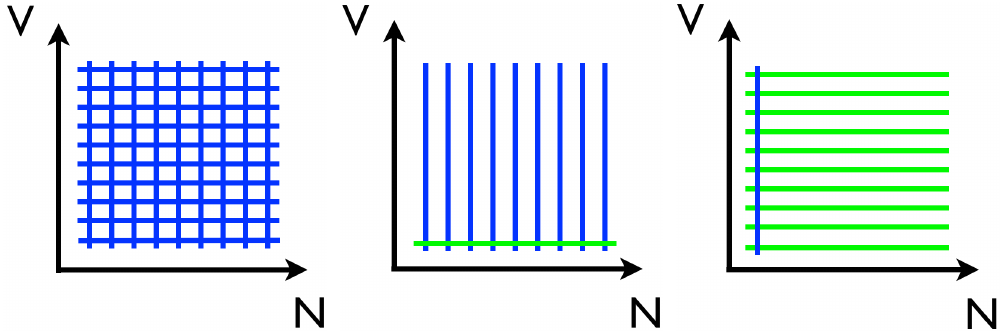}
\par
\caption{\label{reconstruction}Reconstruction schemes of the total osmotic density of state, either by directly computing from a Monte Carlo density of state calculation in the osmotic ensemble (left), or using several WL simulations at fixed adsorption loading $N\e{ads}$, and reconstructing them using a density of state calculation at a given reference volume $V\e{ref}$ (center), or using several WL simulations at fixed volume $V$, and reconstructing them using the density of state of the bare host material, i.e. at $N=0$ (right). }
\end{figure}

Two reconstruction schemes of the total osmotic density of state have been realized (see Fig.~\ref{reconstruction}). The first one consists in computing the one dimensional DOS at fixed adsorption loading $Q(V;N\e{ads})$, and reconstructing the 2D DOS using a density of state calculation at a given reference volume $V\e{ref}$ ($Q(N\e{ads};V\e{ref})$). The second scheme is to calculate the DOS for several fixed volume $V$ as a function of $N\e{ads}$ ($Q(N\e{ads};V)$), and then reconstructing the density of states using the knowledge of a density of state at a given $N\e{ads}$. Both reconstructions leads to the same result within statistical uncertainties. However, the second approach presents a strong advantage since one can use the DOS at $N\e{ads} = 0$. Indeed, taking advantage of the fact that $\Omega\e{WL}(N\e{ads} = 0,V;\mu\e{ads},\sigma) = F\e{host}(V) + \sigma V$, it follows that:
\begin{equation}
\Omega\e{WL} (N\e{ads},V;\mu\e{ads},\sigma) =
 F\e{host}(V) + \sigma V +  \Omega\e{WL}(N\e{ads};\mu\e{ads}, V)
\end{equation}

The use of the known host free energy function as a basis for reconstruction speeds up calculations.  Moreover it allows  to recalculate the 2D free energy profiles for various host free energy profiles without re-running any new calculations since the $F\e{host}$ term is not included directly in the simulations, but only used in post-processing data to reconstruct free energy surface. This is of great help in investigating the effect of the host material's properties on adsorption-induced breathing.

\section{Results and Discussion}

\subsection{Simple MIL-53 Model}

\begin{figure}[ht]
  \begin{center}
    \includegraphics[width=8cm]{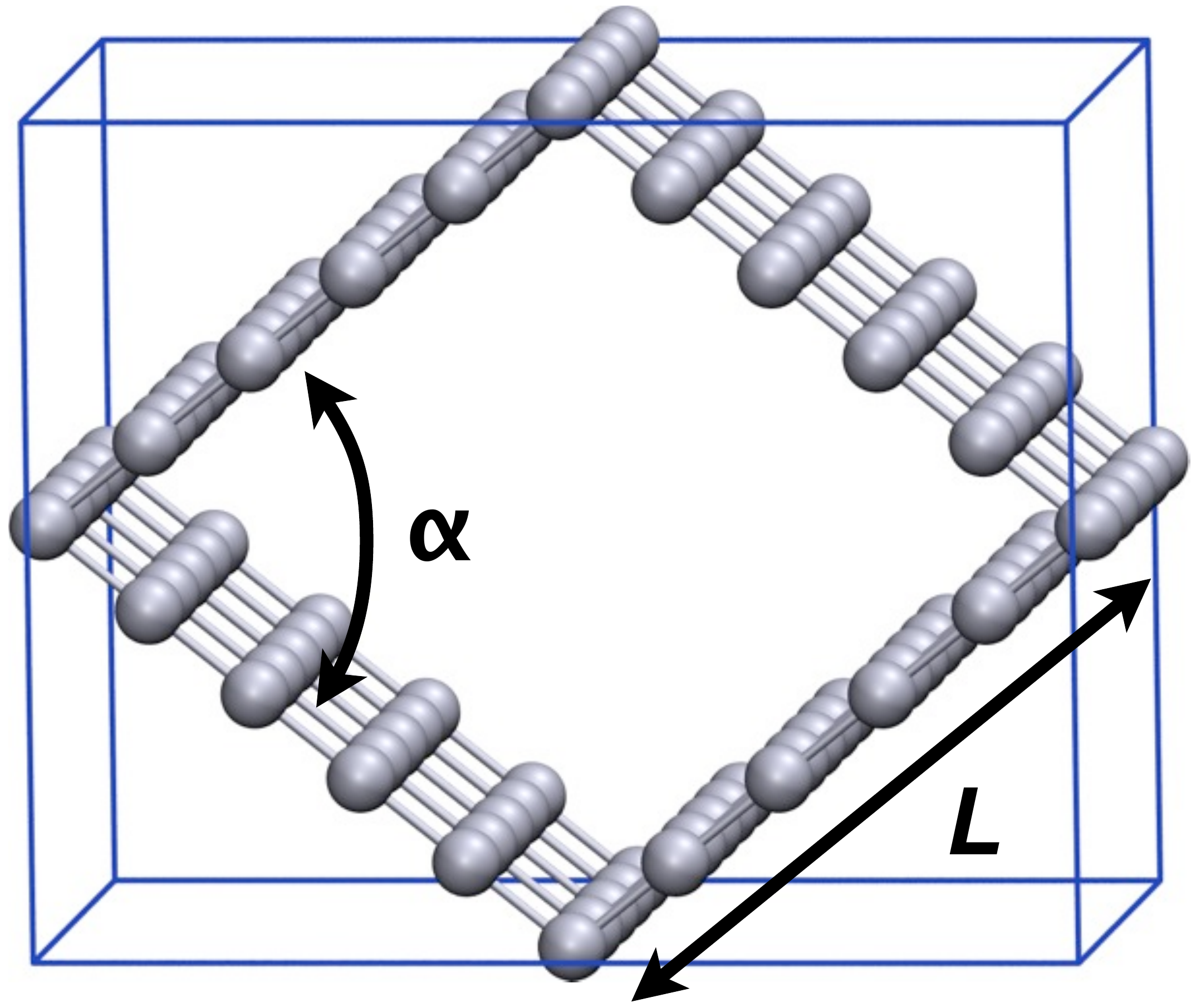}
    \caption{\label{snap}Snapshot of a unit cell of our simplified MIL-53 model.}
  \end{center}
\end{figure}

In order to focus our study on the efficiency of various molecular simulation methods for soft porous crystals, we have limited the computational effort involved in each individual simulation by working on a very simple model mimicking the geometrical and physical properties of the MIL-53 family of ``breathing'' materials. The model, depicted in Fig~\ref{snap},  is made of unidimensional diamond-shaped pores separated by walls of Lennard-Jones particles. The corner to corner distance is denoted by $L$, and the angle between two walls is $\alpha$; the channels are parallel to the $z$ axis. Both values are chosen to mimic the experimental cell parameters of the np and lp phase of the MIL-53(Al): $L$ is taken as {10~\AA} and $\alpha$ is either 42.6\textdegree{} (np phase) or 75.1\textdegree{} (lp phase). The simulation cell in the perpendicular dimension is fixed to  {10~\AA}.  All simulations were then performed on a $2\times 1 \times 2$ supercell with periodic boundaries conditions. 

Wall-fluid and fluid-fluid interactions were both described by 12--6 shifted Lennard-Jones potentials, with a spherical cutoff at a distance of 7.5~\AA. Fluid-fluid interaction parameters are taken as $\sigma\e{ff} = 3.5$~\AA{} and $\epsilon\e{ff} = 150$~K. The wall-fluid interactions were tuned so that the adsorption enthalpy in the np phase be more favorable than in the lp phase, which is the \emph{sine qua non} condition for breathing.\cite{Cou2008} The wall-fluid Lennard-Jones parameters were varied in the range of $2.8\text{~\AA} \le \sigma\e{wf} \le 4.0$~\AA{}, and $0 \le \epsilon\e{wf} \le 300$~K. MC simulations in the NVT ensemble with a single adsorbate molecule were performed to calculate np and lp adsorption enthalpies. Results (plotted in Fig.~S1\footnote{See Supplementary Material Document No. \_\_\_\_\_\_\_\_\_ for plots of additional adsorption properties of the model. For information on Supplementary Material, see http://www.aip.org/pubservs/epaps.html}) show that the adsorption enthalpy difference is maximum for $\sigma\e{wf} = 3.4$~\AA{}. We chose a $\epsilon\e{wf}$ value of $140$~K to get reasonable adsorption at pressures comparable to experiment on the real material. 

\begin{figure}[ht]
  \begin{center}
    \includegraphics[width=8cm]{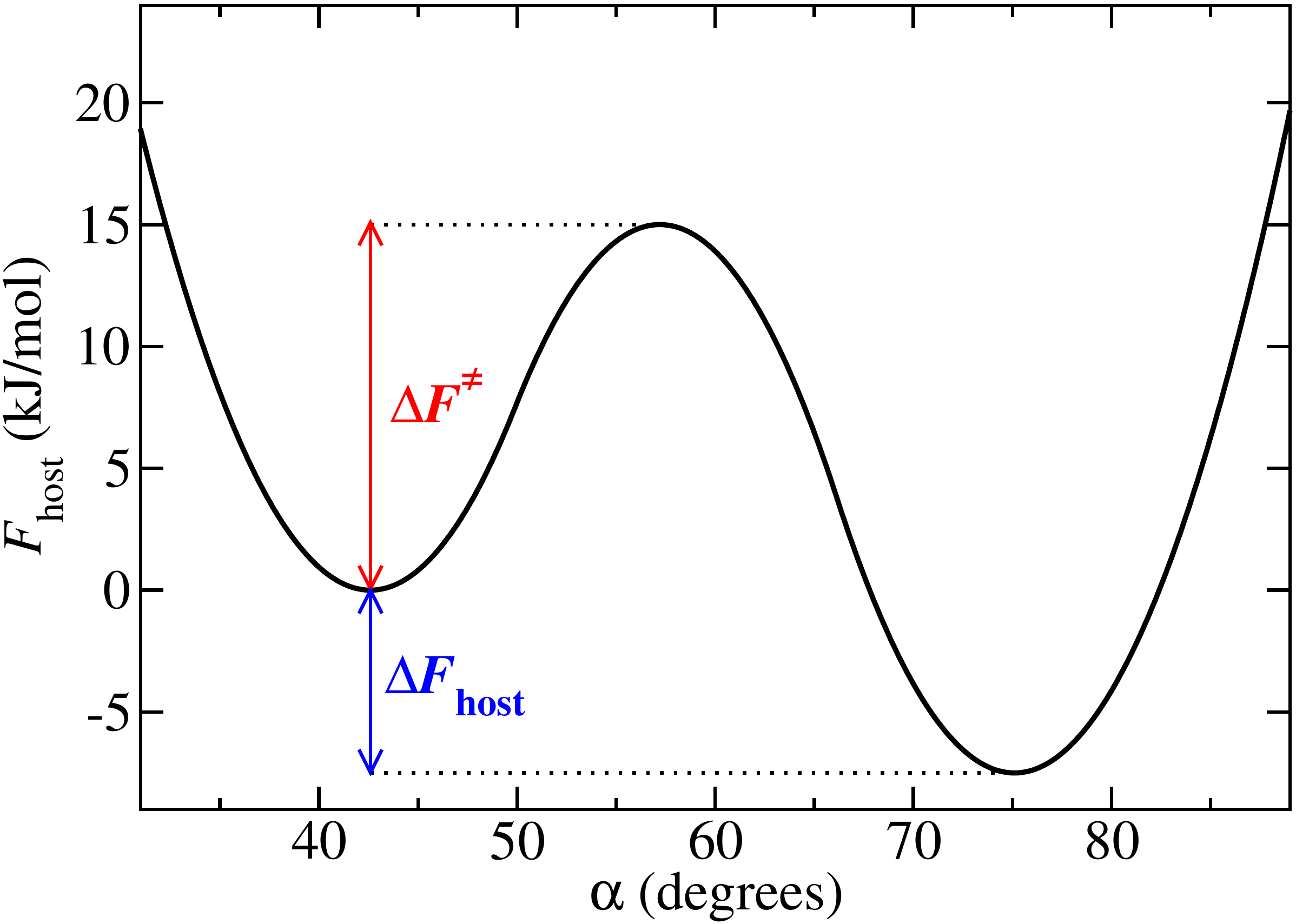}
    \caption{\label{fhost}Free energy profile of the host as a function of its opening $\alpha$, at $T = 300$~K.}
  \end{center}
\end{figure}

A free energy term as a function of the opening $\alpha$ was added in order to account for the bistable nature of the material's framework. The free energy landscape $F\e{host}(\alpha)$ was defined as a smooth biparabolic potential, as shown in Fig.~\ref{fhost}. The free energy difference between np and lp structures was taken from Ref.~\citenum{Cou2008} and fixed at 7.5~kJ.mol$^{-1}$ at 300~K. Unless otherwise specified, the free energy barrier was arbitrary chosen as 15~kJ.mol$^{-1}$.  This simple system is expected to behave only qualitatively as the real material it is inspired from, rather than reproducing any experimental data. This model does not include the rotation of organic linkers for example,  and the only internal degree of freedom of the host considered at this stage is the opening angle. 

\subsection{Grand Canonical Monte Carlo simulations}

\begin{figure}[ht]
  \begin{center}
    \includegraphics[width=8cm]{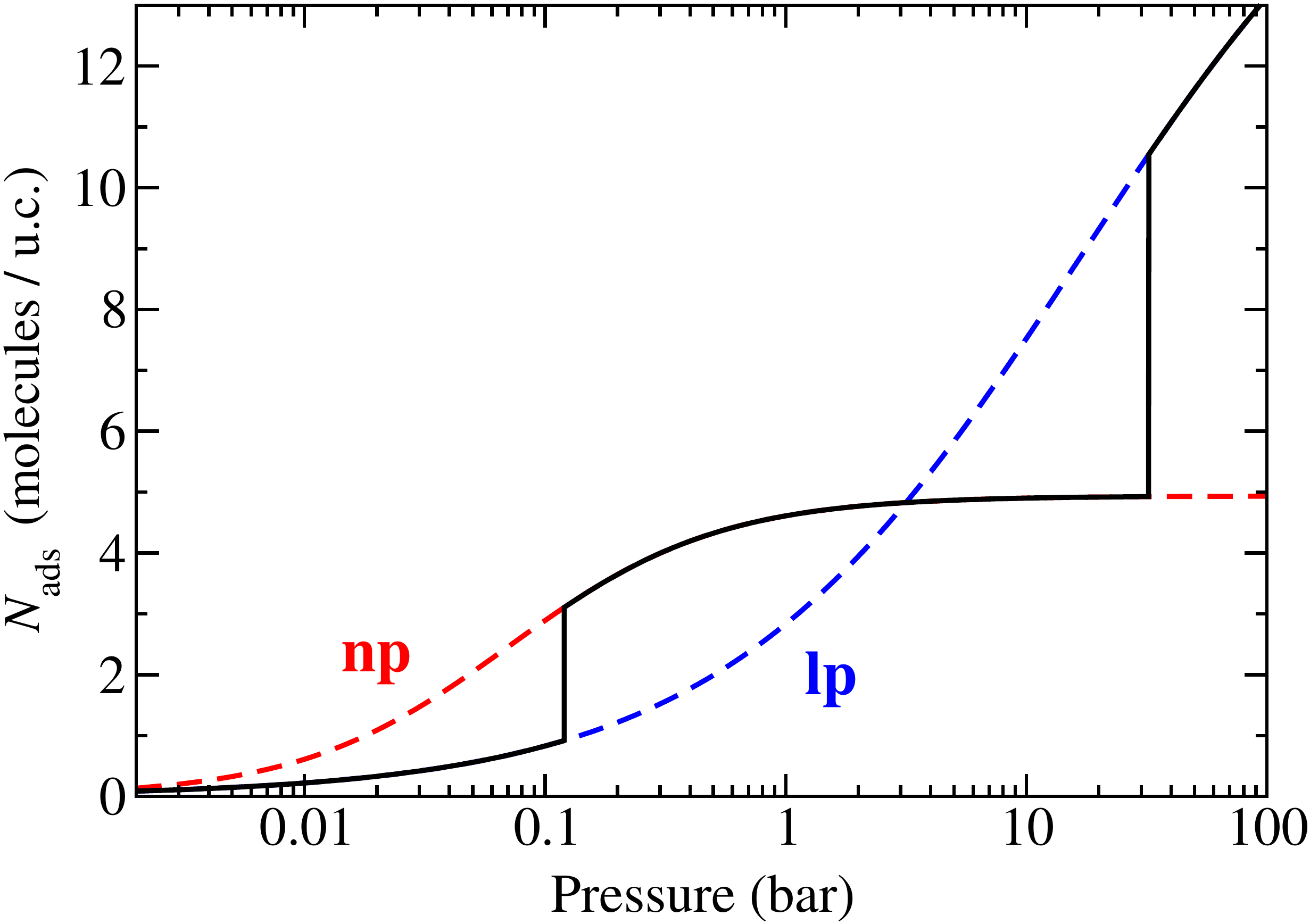}
    \caption{\label{iso}Adsorption isotherms computed at $T=300$~K with GCMC simulations in the np (dashed red) and lp (dashed blue) rigid structures. The full black line is a composite adsorption isotherm, featuring the breathing transitions calculated using an analytical model in the osmotic ensemble.}
  \end{center}
\end{figure}

Before studying the full thermodynamics of the system in the osmotic ensemble, we first performed a study in the grand canonical ensemble. There, the host was considered as rigid and was fixed in either the np phase, or the lp phase, and adsorption isotherms were computed at 300~K using standard GCMC simulations for each phase. The resulting adsorption isotherms are shown as dashed lines in Fig.~\ref{iso}; they depict a behavior typical of the experimental data on MIL-53 breathing. In particular, the Henry constant is significantly higher in the np phase and the maximum adsorbed amount is lower, which are the \emph{sine qua non} conditions for breathing. The adsorbate density maps in the np and lp phases have been analyzed (see Fig.~S2). The lp phase features two different types of adsorption site. 
The osmotic free energy difference between the lp and np phases can be computed by integrating the GCMC adsorption isotherms:\cite{Cou2008,Coh2009}
\begin{equation}
\Delta \Omega (\mu\e{ads} ,\sigma) = \Delta F\ex{host}\e{np-lp} + \sigma \Delta V\e{np-lp} - RT \int_0^P \frac{ \Delta\e{np-lp} N\e{ads}(P')}{P'} dP'
\end{equation}

Finally, using this thermodynamic model, we can produce equilibrium isotherms that feature vertical steps corresponding to lp$\rightarrow$np and np$\rightarrow$lp structural transitions, as depicted on Fig.~\ref{iso} (for $T=300$~K). The respective transition pressures calculated in this approximation are 0.063 and 46.2~bar, respectively. Adsorption in this simple model does trigger breathing at 300~K demonstrating its ability to capture at the microscopic level the features of a breathing framework. We can thus reconstruct a full thermodynamic picture of the system from GCMC isotherms, although one needs to be mindful of the fact this formulation remains incomplete. Indeed, the simulations were performed in the grand canonical ensemble, missing all effects of flexibility and local deformation of the framework. Combining them only gives a representation in a sub-osmotic ensemble, that only takes into account two possible sets of cell parameters corresponding to those of the empty np and lp phases. Explicitly considering the host flexibility is paramount when describing the adsorption-induced flexibility of soft porous crystals.

\subsection{Direct Simulation in the Osmotic Ensemble}

In order to evaluate the convergence and efficiency of direct Monte Carlo simulations for highly flexible materials, we performed a MC simulation of adsorption in our MIL-53 model material in the osmotic ensemble. To do so, the simulation included the standard GCMC moves (insertion, deletion and translation) as well as moves involving changes of the unit cell parameters. Taking into account the constraints due to the nature of the material, the length $L$ of the model was considered fixed (consistent with the rather rigid nature of the organic linkers in the real material) and the volume change steps thus only affected the opening angle, $\alpha$, of the material, homogeneously for all pores in the simulated supercell. ``Adsorption'' and ``desorption'' isotherms were then computed, which corresponded to a series of osmotic ensemble simulations at increasing (resp. decreasing) chemical potential $\mu$, each simulation starting from the final configuration of a previous run at the immediately lower (resp. higher) value of $\mu$. All simulations were started from the lp phase: the unit cell was initially empty for adsorption, or full preloaded with adsorbate for desorption calculations. Each simulation was carried for 100 millions MC steps, with chemical potentials corresponding to a total pressure ranging from 0.01 to 1000~bar.

Several simulations were performed using different MC moves set up, more specifically, different insertion and structure change moves probabilities were tested, ranging from 10\% to 20\%. The combination of 20\% insertion/deletion move, 20\% structure change move and 60\% guest molecules translation yields the more transitions between the large and the narrow pores forms. The acceptance probabilities of insertion/deletion moves ranges from 10\% at low pressure to around 1\% at high guest loading, while the acceptance of structure change MC move consistently stay around 5\%, although most of the accepted moves are of a low amplitude.

The number of transition between the closed and the open form is strikingly low for a vast majority of the simulation which presents np-lp switches only at very specific pressure range as depicted in Fig.~\ref{xlp}. At $P=0.02$~bar, one only witness a low number of very short transition to the np phase. At $P = 0.04$~bar, about 20 np$\leftrightarrow$lp transitions are noticed during a simulation of 100 million MC steps. At $P = 0.08$~bar, only one transition occurs from the lp to the np after a few millions steps, and the system is stuck in the closed form until the end of the simulation run (i.e. $10^8$ steps). During desorption, no significant volume change is observed till 0.64 bar where there is only one lp$\rightarrow$np transition during 100 million steps. As pressure lowers, the system behaves exactly as it does during adsorption, i.e. presenting multiple transitions.

Upon adsorption, the first breathing transition np$\rightarrow$lp is obviously observed although the low number of large fluctuation gives doubt about the accurate convergence toward the true thermodynamical equilibrium. The second transition has never been observed in any simulations we have tried. This transition would require a complex change in configurational space involving changes in particles coordinates, guest quantity, and cell parameters. The opening change steps only scale the adsorbate position according to the new cell parameters, and fail to take into account such necessary configurational changes. The reason the first transition is observed is due to the low adsorption loading at which the transition occurs, which renders the configurational change much less substantial, contrary to the second transition that occurs at higher pressure, hence higher loading. 

During desorption simulation, results are essentially the same and only the low pressure transition is spotted, although the pictures highly hinge on the initial condition. A simulation starting from an initial configuration in the lp phase is forced to be in the expected state at high pressure, but one does not witness the closing of the structure until the pressure is orders of magnitude below the equilibrium transition. Such an hysteresis is only due to the incapacity of reaching a statistically significant thermodynamic equilibrium using a direct simulation in the osmotic ensemble. 

The inability of this raw implementation to reproduce the second transition of the breathing phenomenon is in tune with previous published results. As it was expected, the Monte Carlo scheme we used to simulate the system in the osmotic ensemble is equivalent to the NPT Molecular Dynamics used by Maurin and his coworkers.\cite{Gho2010} It appears that there is an inherent obstacle to the simulation of this system in the osmotic ensemble, and one would need to recourse to complex MC moves combining guest molecules insertion/deletion, adsorbate coordinates translation, and host framework structure change, with no guaranty of convergence within an accessible computational time without resorting to complex statistical bias. 

\subsection{Wang--Landau Free Energy Profiles}

To overcome the convergence difficulties in direct osmotic simulations, we implemented the non-Boltzmann sampling method described in the ``Theory'' section, based on the Wang--Landau algorithm. We have thus calculated the full density of states in the $(V,N\e{ads})$ space, from which the full thermodynamic behavior of the system can be deduced. Given the geometry of the material studied, sampling was performed using the diamond-shaped channel angle, $\alpha$, as an order parameter instead of the unit cell volume $V$. 
Thus, a series of Wang--Landau calculations were performed at fixed values of $\alpha$, going from 35\textdegree{} to 86\textdegree{}, to obtain free energy profiles $\Omega\e{WL}(N\e{ads};\mu\e{ads},\alpha)$. Each of these 52 Wang--Landau calculations performed 23 iterations, with the last (and longest) iteration featuring from 250 millions to 1 billion MC steps. From these, the full 2D density of states was calculated using the reconstruction scheme described previously. The uncertainty on the Landau free energy is then expected to be of the order of $10\ex{-3}$J.mol$\ex{-1}$. In the case of a 
simulation of a more complex and realistic system, the number of Wang Landau cycles could be reduced while still getting reasonable accuracy.

\begin{figure}[ht]
  \begin{center}
\includegraphics[width=0.85\textwidth]{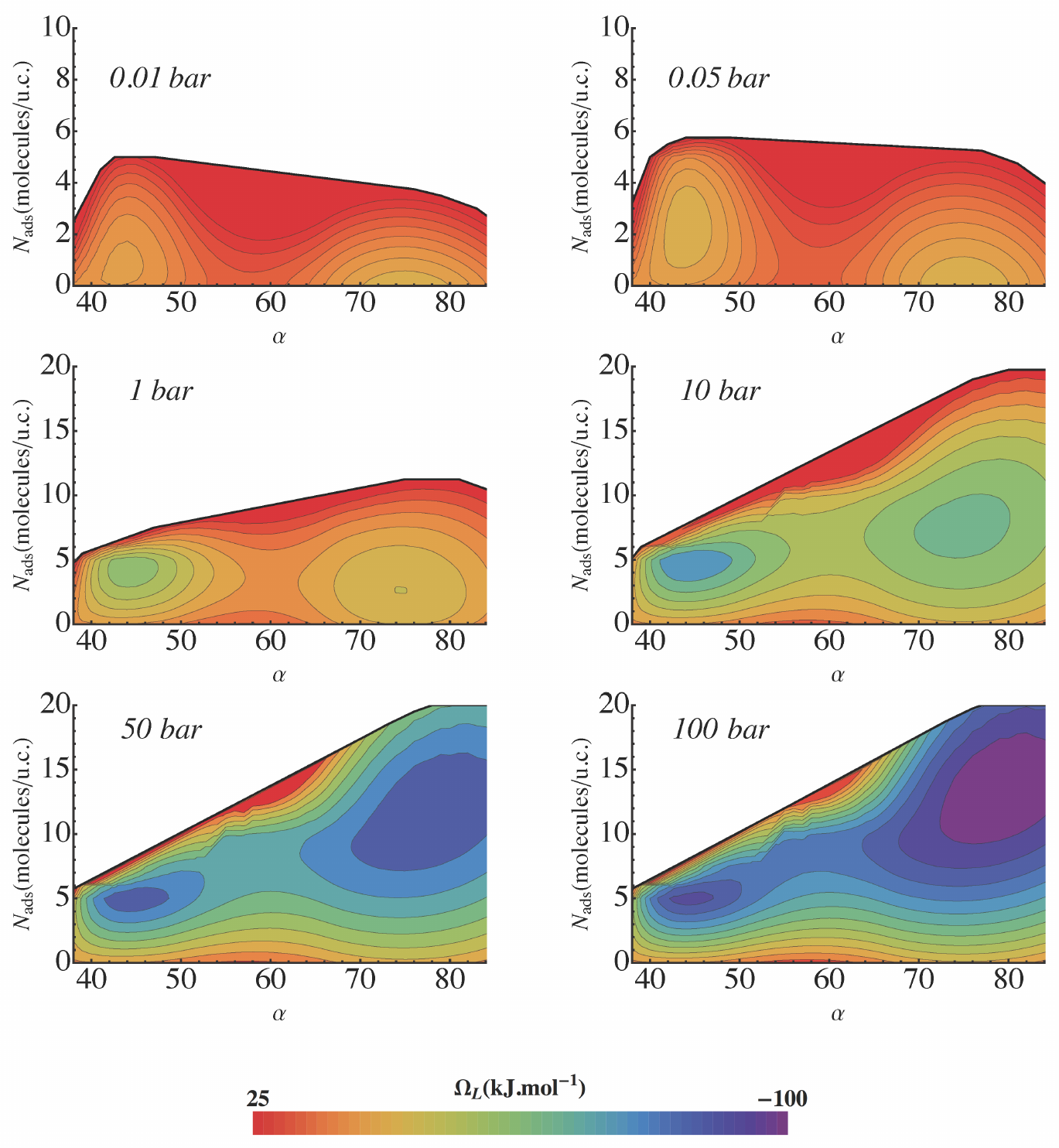}
\caption{\label{dos}Evolution of the osmotic free energy landscape in $(\alpha,N\e{ads})$ space as external gas pressure increases at 300~K: at 0.01~bar (top left), 0.05~bar(top right), 1~bar (middle left), 10~bar (middle right), 50~bar (bottom left) and 100~bar (bottom right).}
  \end{center}
\end{figure}

\begin{figure}[ht]
  \begin{center}
    \includegraphics[width=8cm]{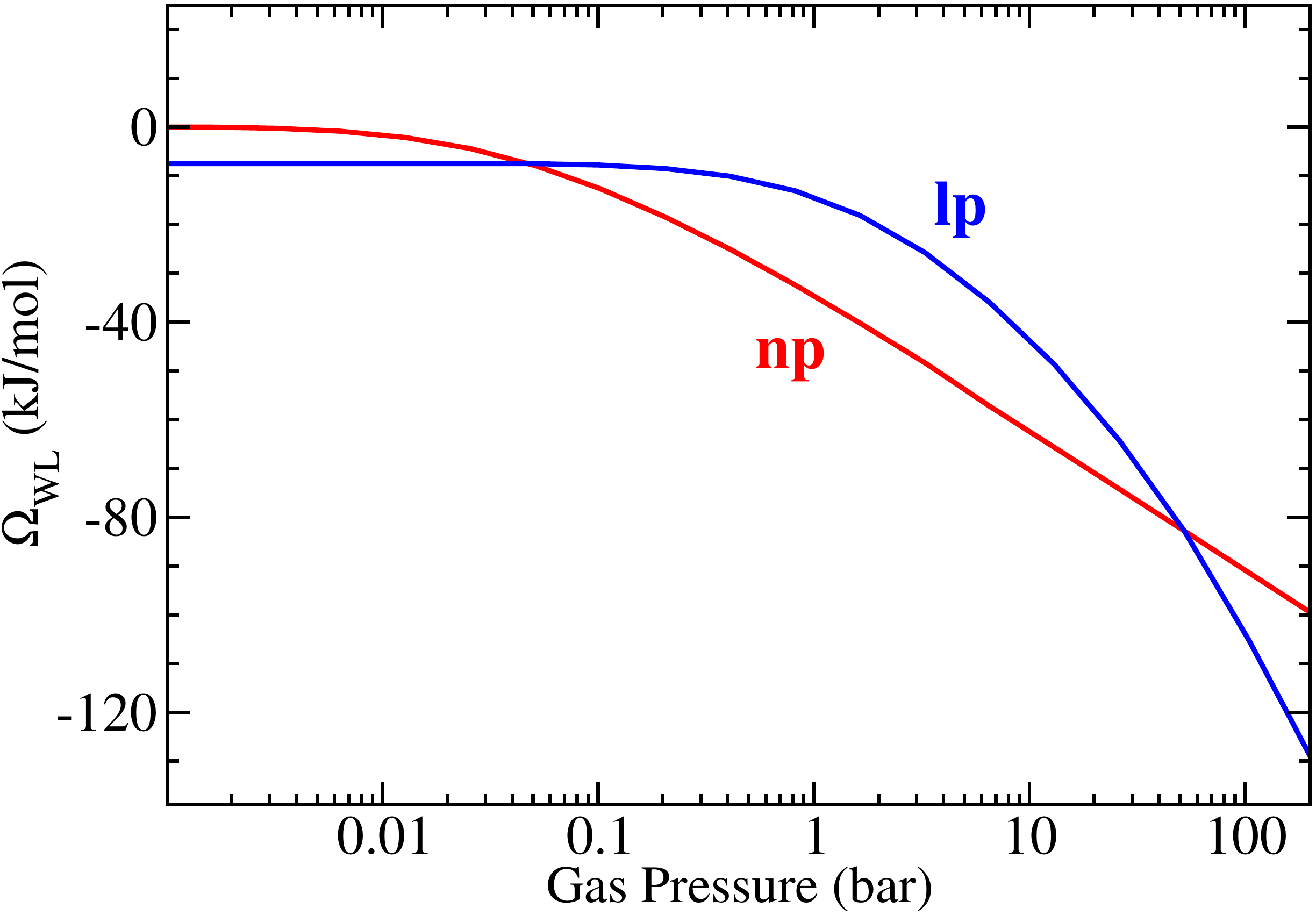}
    \caption{\label{fnplp}Landau free energy of the local minima of the free energy landscapes (Fig.~\ref{dos}), which correspond to the np and lp metastable phases, as a function of external gas pressure at 300~K.}
  \end{center}
\end{figure}

We report in Fig.~\ref{dos} the calculated osmotic-ensemble free energy landscapes as a 2D-function of the opening, $\alpha$, and the number of adsorbed guest molecules, $N\e{ads}$, for a set of increasing chemical potentials $\mu$ corresponding to external gas pressures in the range 0.001 to 100~bar. The landscape is characterized by the presence of two wells corresponding to metastable forms of the host material. A saddle point is found between the two minima that corresponds to the optimal pathways between the two phases. These broad features are direct consequences the bistable nature of the host, taken into account by the biparabolic nature of free energy profile (Fig.~\ref{fhost}). At low pressure, this host free energy dominates the 2D landscape, and favors the lp form as the most stable at 300~K. At slightly higher gas pressure, the minimum of the np region shifts and now corresponds to a non-zero adsorbate loading, while that of the lp phase does not. This is in keeping with the np phase's higher affinity for the guest, as established by the comparison of their Henry constants. 
In addition to 2D free energy surfaces, the evolution of lp and np free energy, i.e. the free energy of the local minima in the free energy surfaces, as a function of gas pressure is shown in Fig.~\ref{fnplp}.

It can there be seen that the adsorption-induced stabilization provided by this adsorption leads the np phase to become more stable than the lp one, above $P=0.051$~bar. This crossing corresponds to the equilibrium of the first breathing transition lp$\rightarrow$np; this pressure is in line with the results of both the direct osmotic simulations and the grand canonical thermodynamic integration. Finally, with further increase in the external gas pressure, the np form's loading reaches saturation, while adsorption in the lp phase now increases markedly, stabilizing this latter. The free energy difference between both phases eventually start to decrease at approximately 3~bar, and the second breathing transition np$\rightarrow$lp is observed at an equilibrium pressure of 50.9~bar.  

\begin{figure}[ht]
  \begin{center}
    \includegraphics[width=8cm]{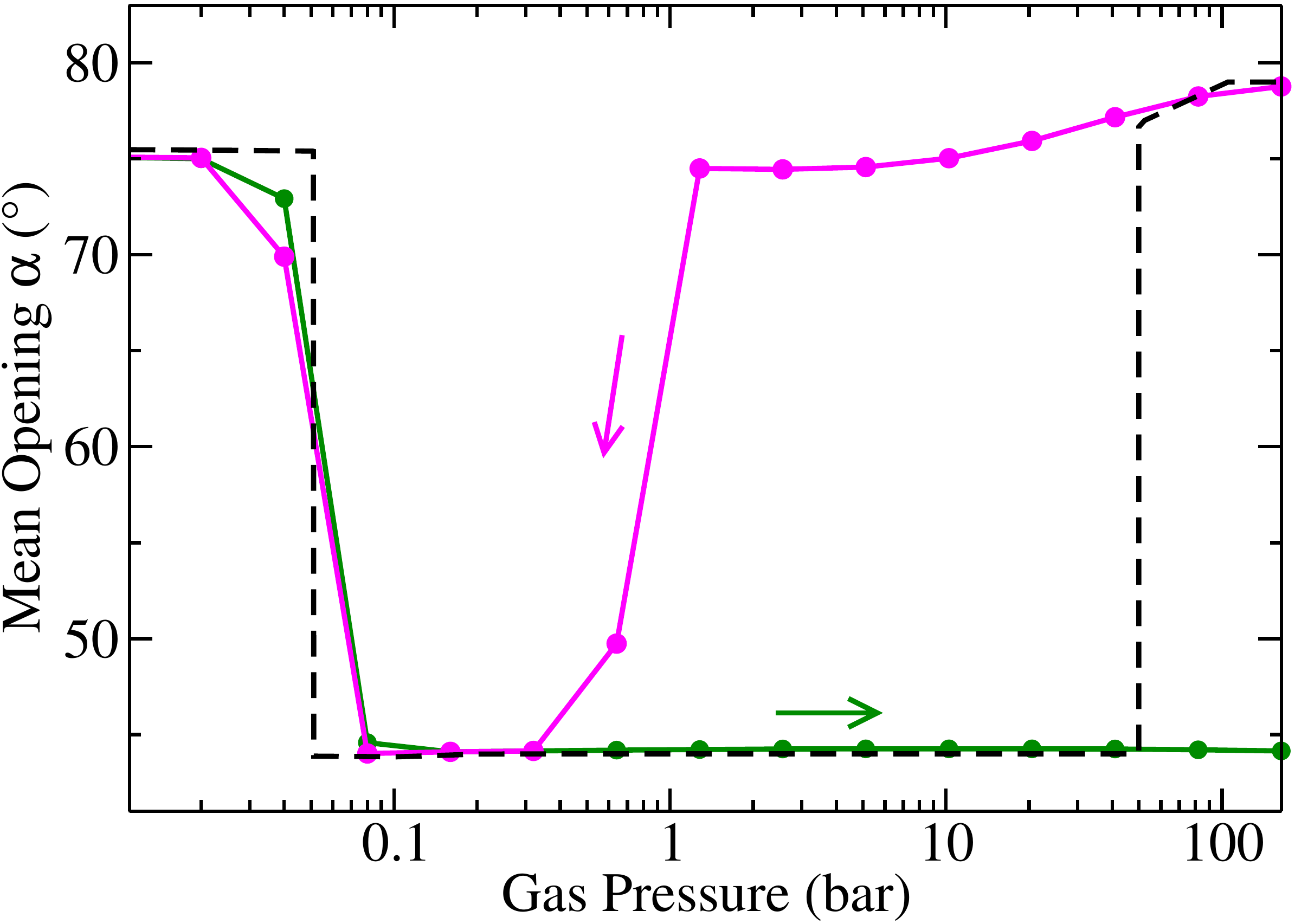}
    \caption{\label{xlp}Mean opening angle as a function of external gas pressure during osmotic Monte Carlo simulations of adsorption (green) and desorption (magenta), at $T=300$~K. The behavior for the system at thermodynamic equilibrium is displayed as a dashed black line (DOS calculation).}
  \end{center}
\end{figure}

\begin{figure}[ht]
  \begin{center}
    \includegraphics[width=8cm]{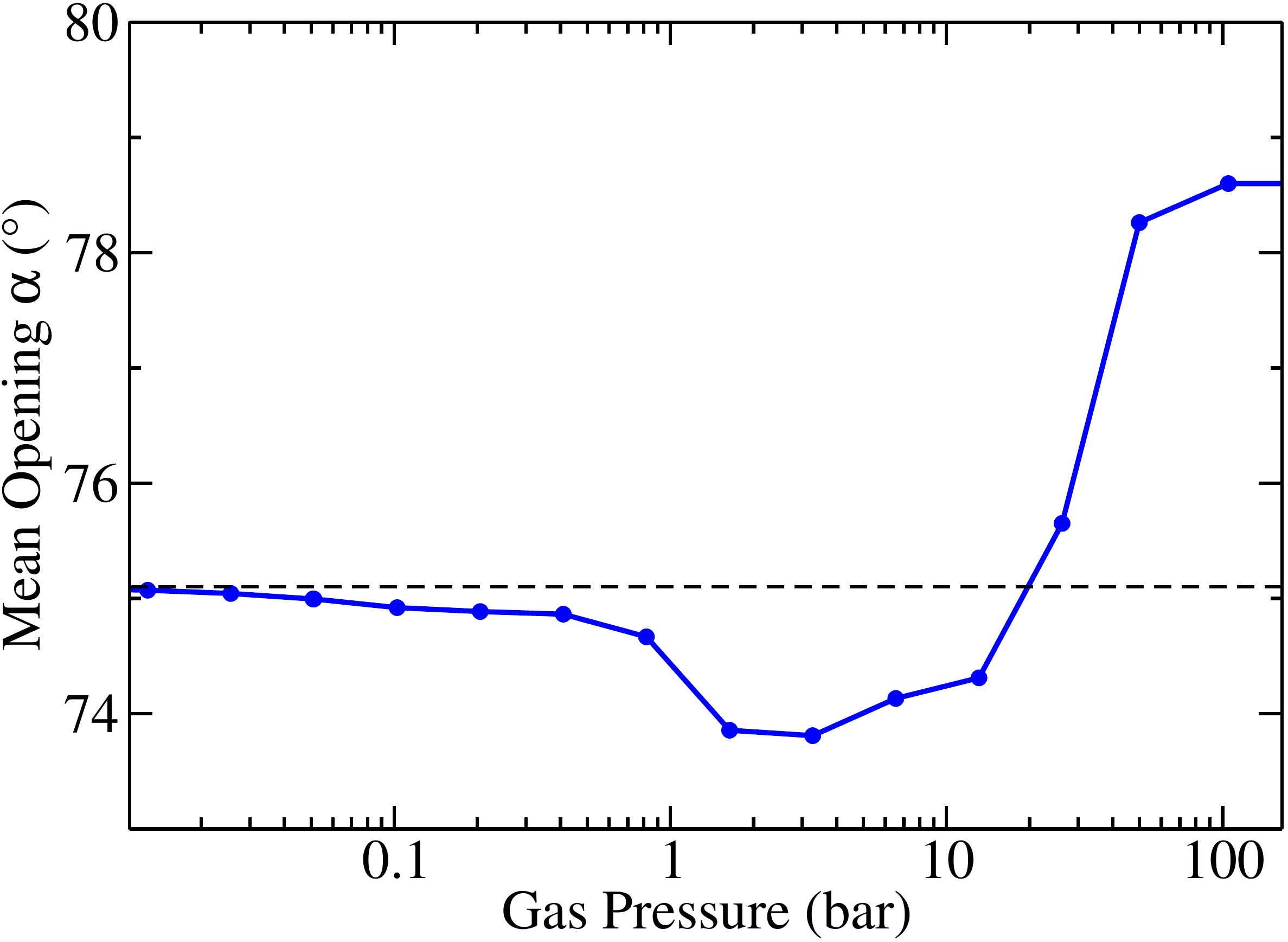}
    \caption{\label{alphalp}Evolution of the opening angle of the lp phase as a function of gas pressure computed from the osmotic density of state calculation}
  \end{center}
\end{figure}

The evolution of the opening angle extracted from the minimum of the free energy landscape as a function of the gas pressure is shown in Fig. \ref{xlp} and compared to direct osmotic simulation results. The first transition matches with direct osmotic simulation results. The second transition clearly stands out in the DOS formulation while, in the direct simulation in the osmotic ensemble, it is only observed during desorption simulation and at a pressure widely shifted from the equilibrium value (see Fig. \ref{xlp}). Our free energy calculation allows to overcome the convergence breakdown in direct osmotic simulation and clearly reveals both transition.
These simulations also agrees on the local contraction/expansion of the material  upon adsorption in each structural phases. The opening angle of the lp phase is plotted on the whole gas pressure range in Fig. \ref{alphalp} disregarding whether the lp phase is a stable or metastable state. Starting from the equilibrium angle of the empty material, the opening angle slightly decreases upon gas adsorption, reaches a minimum and significantly increases at higher gas pressure. Such contraction-dilation phenomenon upon adsorption is a classical behavior observed in many porous material.\cite{Tvardovskiy} It is worth mentioning that such local \emph{flexibility} is linked to the elastic constant of the phase but is not intrinsically related to the large flexibility (or structural transitions) observed in this bi-stable material.\cite{Nei2011}

We then compared the transition pressures obtained with our new approach to the values predicted by our analytical model applied to GCMC isotherms. For both transitions, a discrepancy is observed: 0.051~bar (Wang--Landau) vs. 0.062~bar (GCMC) for the first transition, and 50.9~bar vs. 46~bar for the second one. These differences can be ascribed to adsorption induced elastic deformations of the lp and np phase, which are accounted for in the osmotic ensemble (see Fig.~\ref{alphalp}) but not in simple GCMC adsorption isotherms. Indeed, we calculated the Wang--Landau landscapes for a series of modified host free energy functions, in which both the lp and np phase were rendered less compliant (their elastic constant $K$ was multiplied by a factor ranging from $1.8$ to $1350$).\footnote{It is worth noting that, with the two-step methodology proposed in this paper for calculating these 2D free energy surfaces, we did not have to perform any additional molecular simulations to obtain results with modified host free energy profiles.} These simulations showed an evolution of the transition pressures towards the values obtained from the model applied to GCMC isotherms (see Fig.~S3). The two methods are thus fully consistent.

\begin{figure}[ht]
  \begin{center}
    \includegraphics[width=10cm]{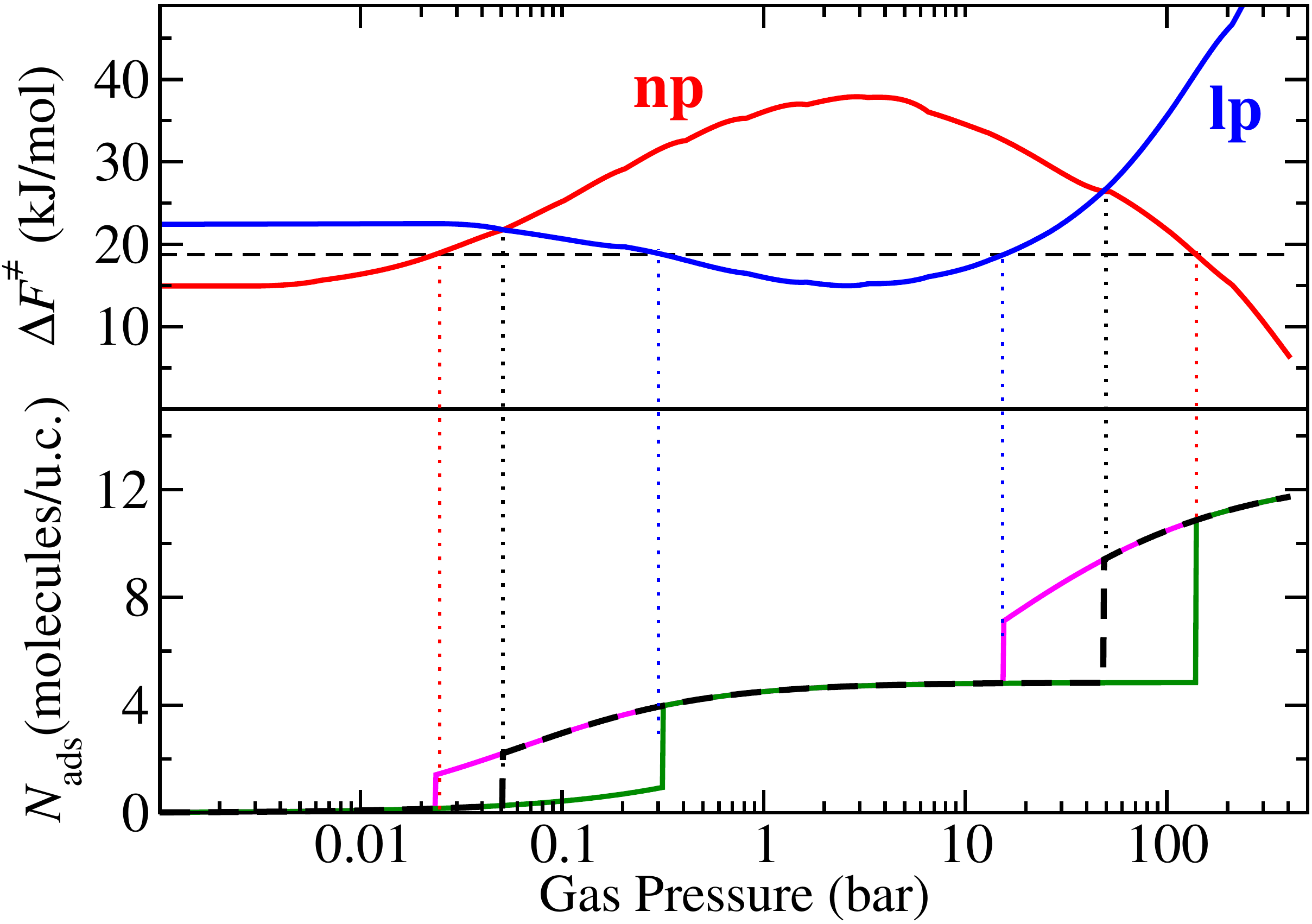}
    \caption{\label{hyst}Top panel: free energy barrier for the np $\rightarrow$ lp and lp $\rightarrow$ np breathing transitions (in red and blue respectively), as a function of external gas pressure, compared to a threshold of 18.75~kJ.mol$^{-1}$ (dashed black line; see text for details). Lower panel: computed adsorption (green) and desorption (orange) isotherms showing hysteresis loops; the isotherm at thermodynamic equilibrium is shown as a dashed black line.}
  \end{center}
\end{figure}

From the full free energy landscape, we have thus analyzed the relative stability of the metastable states, and found it to be different from the results one would have obtained from either direct osmotic molecular simulation (which fail to give equilibrium transition pressures) and GCMC isotherms (which fail to account for local framework deformations). Moreover, we can also look at the full picture of the breathing phenomenon in the $(N\e{ads},V)$ space. On the free energy surface, we can determine a ``reaction path'' for the breathing transitions, and in particular locate the transition state of the each transition, which corresponds to the saddle point of the surface. This enables us to estimate the free energy barriers involved in the np$\rightarrow$lp and lp$\rightarrow$np transitions. The existence of this barrier can account for the hysteretic nature of the system, trapping it in a metastable state. Earlier studies have shown that this free energy barrier influences the dynamics of the breathing transition.\cite{Tri2011} We present in Figure~\ref{hyst} the evolution of the free energy barrier for both the closing and opening transitions as a function of external gas pressure, computed from the analysis of the osmotic free energy landscape. In the figure, we compare these barriers to a threshold value of 18.75~kJ.mol$^{-1}$ over which the transition is considered impossible, and under which the transition occurs immediately. A theoretical framework to study this phenomenon is still lacking and this value remains arbitrary (other barrier thresholds were used to compute the corresponding adsorption-desorption hysteresis loop and are shown in Fig.~S4), but it enables us to give a simple interpretation of the hysteresis loop on the basis of free energy barriers involved in the transitions.

On this basis, we describe here a simple adsorption-desorption cycle, starting in the empty lp phase. Upon adsorption of guest molecules, the free energy barrier of the lp$\rightarrow$np transition lowers until it reaches the critical value at which the system switches to the now much more stable np phase. This necessarily happens at a pressure higher than the equilibrium transition, as expected. As pressure further rises, the np phase eventually ends up saturated in guest molecules and the saddle point (and lp) free energy lower faster than the np minimum, up to a point where the np$\rightarrow$lp free energy barrier meets the threshold. This transition again occurs at higher pressures than the equilibrium between both phases. Then, upon desorption, the free energy barrier of the lp$\rightarrow$np decreases to the point where it gets below the free energy threshold and the material closes itself at a pressure lower than the equilibrium, creating a hysteresis loop. The same phenomenon appears for the last reopening transition, and we see that adsorption and desorption branches display different breathing transitions, creating the two hysteresis loops shown in Fig.~\ref{hyst} and fully compatible with experimental results on breathing MOF's. This hysteresis in the adsorption-desorption loop is highly dependent on the free energy activation value chosen for the analysis, and the assumption made that the threshold remains the same for both the opening and closing transitions as well as being independent of the pressure. Regardless of those hypotheses, this density of state formulation provides a qualitative interpretation of the hysteresis nature of the adsorption-desorption in flexible porous materials

\section{Conclusion}

We have developed a new Monte Carlo simulation method for the simulation of Soft Porous Crystals, based on non-Boltzmann sampling in \{guest loading, volume\} space using the Wang--Landau algorithm. This simulation approach can be used to fully characterize the adsorption properties and the material's response to adsorption at thermodynamic equilibrium. In addition, it produces the full free energy landscape of the material as a function of both deformation and loading, and can thus be used to better understand the hysteretic nature of adsorption-induced structural transitions.  We showcased this new method on a simple model of the MIL-53 family of breathing materials, highlighting its advantages compared with the conventional simulation techniques used in the field (Grand Canonical Monte Carlo of rigid structures, and direct simulations in the osmotic ensemble). This method will be applied to atomistic models of real materials using realistic host forcefields in future work.

\vskip 3cm

\section*{Acknowledgments}

The authors thank Fabien Cailliez, Alex Neimark, and Alain Fuchs for discussions and insightful comments. They acknowledge funding from the Agence Nationale de la Recherche under project ``SOFT-CRYSTAB'' (ANR-2010-BLAN-0822).

\clearpage


%

\clearpage

\end{document}